\documentclass[12pt]{iopart}
\usepackage{amsfonts}
\usepackage{iopams}
\usepackage{epsfig}
\usepackage{cite}

\begin{document}

\title{Morphological characterization of shocked porous material}
\author{Aiguo Xu, Guangcai Zhang, X. F. Pan, Ping Zhang, and Jianshi Zhu}
\address{National Key Laboratory of Computational Physics, \\
Institute of Applied Physics and Computational Mathematics, P. O.
Box 8009-26, Beijing 100088, P.R.China} \ead{Xu\_Aiguo@iapcm.ac.cn}
\date{\today }

\begin{abstract}
Morphological measures are introduced to probe the complex procedure
of shock wave reaction on porous material. They characterize the
geometry and topology of the pixelized map of a state variable like
the temperature. Relevance of them to thermodynamical properties of
material is revealed and various experimental conditions are
simulated. Numerical results indicate that, the shock wave reaction
results in a complicated sequence of compressions and rarefactions
in porous material. The increasing rate of the total fractional
white area $A$ roughly gives the velocity $D$ of a
compressive-wave-series. When a velocity $D$ is mentioned, the
corresponding threshold contour-level of the state variable, like
the temperature, should also be stated. When the threshold
contour-level increases, $D$ becomes smaller. The area $A$ increases
parabolically with time $t$ during the initial period. The $A(t)$
curve goes back to be linear in the following three cases: (i) when
the porosity $\delta$ approaches 1, (ii) when the initial shock
becomes stronger, (iii) when the contour-level approaches the
minimum value of the state variable. The area with high-temperature
may continue to increase even after the early compressive-waves have
arrived at the downstream free surface and some rarefactive-waves
have come back into the target body. In the case of energetic
material needing a higher temperature for initiation, a higher
porosity is preferred and the material may be initiated after the
precursory compressive-waves have scanned all the target body. One
may desire the fabrication of a porous body and choose appropriate
shock strength according to what needed is scattered or connected
hot-spots. With the Minkowski measures, the dependence on
experimental conditions is reflected simply by a few coefficients.
They may be used as order parameters to classify the maps of
physical variables in a similar way like thermodynamic phase
transitions.
\end{abstract}

\vspace{2pc}
\submitto{\JPD} \maketitle


\section{Introduction}
A porous material contains voids or tunnels of different shapes and
sizes. Such materials are commonly found in nature and as industrial
materials such as wood, carbon, foams, ceramics, bricks, metals and
explosives. They have also been used in surgical implant design to
fabricate devices to replace or augment soft and hard tissues, etc.
In order to use them effectively, their mechanical and
thermodynamical properties must be understood in relation to their
mesoscopic structures\cite{v2_1,v2_2}.

In this work we focus on porous materials under shock wave reaction.
When a porous material is shocked, the cavities inside the sample
may result in jets and influence its back velocity\cite{n1}. Cavity
nucleation due to tension waves controls the spallation behavior of
the material\cite{n3}. Cavity collapse plays a prominent role in the
initiation of energetic reactions in explosives\cite{B2002}. In this
side, most of previous studies concerned the
Hugoniots\cite{e1,e2,e3,e4,e5,e6,t1,t4} and the equation of
state\cite{eos1,eos2,eos3}. It is known that, under strong shocks,
the porous material is globally in a nonequilibrium state and show
complex dissipative structures. How to describe and pick up
information from such a system is still an open problem. In this
work we introduce the Minkowski functionals to measure the
morphological behaviors of the map of state variable and use them to
probe the procedure of shock wave reaction on porous material.

This study needs also a powerful simulation tool. The molecular
dynamics can discover some atomistic mechanisms of shock-induced
void collapse\cite{Porous7,Yang}, but the spatial and temporal
scales it may cover are far from those comparable with experiments.
To overcome this scale limitation, we resort to a newly developed
mesoscopic particle method, the material-point
method(MPM)\cite{H1964,MPM1,MPM2,MPM3,MPM4,MPM5}. The MPM was
originally introduced in fluid dynamics by Harlow, et al\cite{H1964}
and extended to solid mechanics by Burgess, et al\cite{MPM1}, then
developed by various researchers, including
us\cite{JPCM2007,CTP2008,JPD2008}. The other reason for using the
MPM is related to the severe difficulties of the traditional
Eulerian and Lagrangian methods in treating with shocked porous
materials. The material under investigation is generally highly
distorted during the collapsing of cavities. The Eulerian
description is not convenient for tracking interfaces. When the
Lagrangian formulation is used, the original element mesh becomes
distorted so significantly that the mesh has to be re-zoned to
restore proper shapes of elements. The state fields of mass density,
velocities and stresses must be mapped from the distorted mesh to
the newly generated one. This mapping procedure is not a
straightforward task, and introduces errors. The MPM not only takes
advantages of both the Lagrangian and Eulerian algorithms but makes
it possible to avoid their drawbacks as well.
 At each time step, calculations consist of two parts: a Lagrangian
part and a convective one. Firstly, the computational mesh deforms
with the body, and is used to determine the strain increment, and
the stresses in the sequel. Then, the new position of the
computational mesh is chosen (particularly, it may be the previous
one), and the velocity field is mapped from the particles to the
mesh nodes. Nodal velocities are determined using the equivalence of
momentum calculated for the particles and for the computational
grid.

The following part of the paper is planned as follows. Section 2
briefly reviews the Minkowski descriptions. Section 3 presents the
theoretical model of the material under consideration. Simulation
results are shown and analyzed in section 4. Section 5 makes the
conclusion.

\section{Brief review of morphological characterization}

A variety of techniques can be used to describe the complex spatial
distribution and time evolution of state variables in the shocked
porous material. In this study we concentrate on the set of
statistics known as Minkowski functionals\cite{Minkowski1903}. A
general theorem of integral geometry states that all properties of a
$d$-dimensional convex set (or more generally, a finite union of
convex sets) which satisfy translational invariance and additivity
(called morphological properties) are contained in $d+1$ numerical
values \cite{Hadwiger19561959}. For a pixelized map
$\psi(\mathbf{x})$, we consider the excursion sets of the map,
defined as the set of all map pixels with value of $\psi$
greater than some threshold $\psi_{th}$ (see, e.g., Refs. \cite%
{Weinberg1987,Melott1990}), where $\mathbf{x}$ is the position,
$\psi$ can be a state variable like temperature $T$, density $\rho$
or pressure $P$; $\psi$ can also be the velocity $\mathbf{v}$ or its
components, some specific stress, etc. Then the $d+1$ functionals of
these excursion sets completely describe the morphological
properties of the underlying map $\psi(\mathbf{x})$. In the case of
two or three dimensions, the Minkowski functionals have intuitive
geometric interpretations.

For a two-dimensional map, the three Minkowski functionals
correspond geometrically to the total fractional area $A$ of the
excursion set, the boundary length $L$ of the excursion set per unit
area, and the Euler characteristic $\chi $ per unit area (equivalent
to the topological genus). Such a description has been successfully
used to describe patterns in reaction-diffusion
system\cite{PRE1996}, the cosmic microwave background temperature
fluctuations\cite{astro-ph1997}, and patterns in phase separation of
complex fluids\cite{JChemP2000,Sofonea,PRE2003,PR}, etc.

In this work we probe the shocked porous material via checking the
temperature map $T(\mathbf{x},t)$, where the time $t$ is explicitly
denoted. The maps of other physical variables can be analyzed in a
similar way. When the temperature $T(\mathbf{x})$ is beyond the
threshold value $T_{th}$, the grid node at position $\mathbf{x}$ is
regarded as a white (or hot) vertex, else it is regarded as a black
(or cold) one. For the square lattice, a pixel possesses four
vertices. A region with connected white (hot) or black (cold) pixels
is defined as a white (hot) or black (cold) domain. Two neighboring
white and black domains present a clear interface or boundary. When
we increase the threshold contour-level $T_{th}$ from the lowest
temperature to the highest one in the system, the white area $A$
will decrease from $1$ to $0$; the boundary length $L$ first
increases from $0$, then arrives at a maximum value, finally
decreases to $0$ again. There are several ways to define the Euler
characteristic $\chi $. Two simplest ones are
\begin{equation}
\chi =N_{W}-N_{B}\mathtt{,}  \label{Mink1}
\end{equation}%
or
\begin{equation}
\chi =\frac{N_{W}-N_{B}}{N}\mathtt{,}  \label{Mink2}
\end{equation}%
where $N_{W}$ ($N_{B}$) is the number of connected white (black) domains, $%
N $ is the total number of pixels. The only difference of the two
definitions is that the first keeps $\chi $ an integer. In contrast to the
white area $A$ and boundary length $L$, the Euler characteristic $\chi $
describes the connectivity of the domains in the lattice. It describes the
pattern in a purely topological way, i.e., without referring to any kind of
metric. It is negative (positive) if many disconnected black (white) regions
dominate the image. A vanishing Euler characteristic indicates a highly
connected structure with equal amount of black and white domains.
Specifically, for the definition (\ref{Mink1}), the integer $\chi $ equals $%
-1$ when one has a black drop in a large white lattice, and $+1$ vice versa,
since the surrounding white (black) region does conventionally not count. In
this paper, we use the second definition without making any ambiguity. The
ratio
\begin{equation}
\kappa =\frac{N_{W}-N_{B}}{NL}  \label{Mink3}
\end{equation}%
describes the mean curvature of the boundary line separating black and white
domains. Despite having global meaning, the Euler characteristic $\chi $ can
be calculated in a local way using the additivity relation\cite{PRE1996}.

\begin{figure}[tbp]
\caption{(in JPG format) Configurations with temperature contours.
$\protect\delta=2$ and $v_{init}=1000$m/s. From left to right,
t=500ns,
1500ns, 2000ns, and 2500ns, respectively. The length unit here is 10 $%
\protect\mu $m. }
\end{figure}

\begin{figure}[tbp]
\centerline{\epsfig{file= 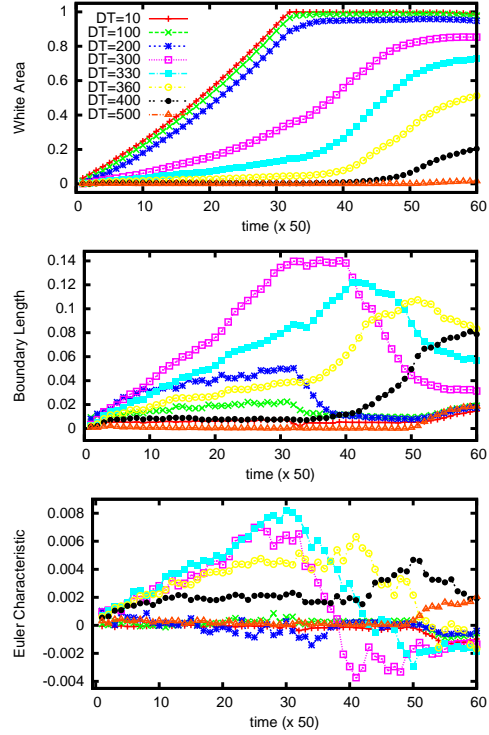, bbllx=103 pt, bblly=99
pt,bburx=509 pt,bbury=701 pt, width=0.42\textwidth,clip=}}
\caption{(Color online) Minkowski measures for the procedure shown in Fig.1.
The contour levels of the temperature increment are shown in the legend.}
\end{figure}

\section{Theoretical model of the material}

In this study the material is assumed to follow an associative von Mises
plasticity model with linear kinematic and isotropic hardening\cite{CModel}.
Introducing a linear isotropic elastic relation, the volumetric plastic
strain is zero, leading to a deviatoric-volumetric decoupling. So, it is
convenient to split the stress and strain tensors, $\boldsymbol{\sigma }$
and $\boldsymbol{\varepsilon }$, as
\begin{eqnarray}
\boldsymbol{\sigma } &=&\mathbf{s}-P\mathbf{I},P=-\frac{1}{3}\verb|Tr|(%
\boldsymbol{\sigma })\mathtt{,}  \label{PMe1} \\
\boldsymbol{\varepsilon } &=&\mathbf{e}+\frac{1}{3}\theta \mathbf{I},\theta =%
\frac{1}{3}\verb|Tr|(\boldsymbol{\varepsilon })\mathtt{,}  \label{PMe2}
\end{eqnarray}%
where $P$ is the pressure scalar, $\mathbf{s}$ the deviatoric stress tensor,
and $\mathbf{e}$ the deviatoric strain. The strain $\mathbf{e}$ is generally
decomposed as $\mathbf{e}=\mathbf{e}^{e}+\mathbf{e}^{p}$, where $\mathbf{e}%
^{e}$ and $\mathbf{e}^{p}$ are the traceless elastic and plastic components,
respectively. The material shows a linear elastic response until the von
Mises yield criterion,
\begin{equation}
\sqrt{\frac{3}{2}}\left\Vert \mathbf{s}\right\Vert =\sigma _{Y}\mathtt{,}
\label{PM1}
\end{equation}%
is reached, where $\sigma _{Y}$ is the plastic yield stress. The yield $%
\sigma _{Y}$ increases linearly with the second invariant of the plastic
strain tensor $\mathbf{e}^{p}$, i.e.,
\begin{equation}
\sigma _{Y}=\sigma _{Y0}+E_{\tan }\left\Vert \mathbf{e}^{p}\right\Vert
\mathtt{,}  \label{PM4}
\end{equation}%
where $\sigma _{Y0}$ is the initial yield stress and $E_{\tan }$ the
tangential module. The deviatoric stress $\mathbf{s}$ is calculated by
\begin{equation}
\mathbf{s}=\frac{E}{1+\nu }\mathbf{e}^{e}\mathtt{,}  \label{s}
\end{equation}%
where $E$ is the Yang's module and $\nu $ the Poisson's ratio. Denote the
initial material density and sound speed by $\rho _{0}$ and $c_{0}$,
respectively. The shock speed $U_{s}$ and the particle speed $U_{p}$ after
the shock follows a linear relation, $U_{s}=c_{0}+\lambda U_{p}$, where $%
\lambda $ is a characteristic coefficient of material. The pressure $P$ is
calculated by using the Mie-Gr\"{u}neissen state of equation which can be
written as
\begin{equation}
P-P_{H}=\frac{\gamma (V)}{V}[E-E_{H}(V_{H})]  \label{eq-eos}
\end{equation}%
In Eq.(\ref{eq-eos}), $P_{H}$, $V_{H}$ and $E_{H}$ are pressure,
specific volume and energy on the
Rankine-Hugoniot curve, respectively. The relation between $P_{H}$ and $%
V_{H} $ can be estimated by experiment and can be written as
\begin{equation}
P_{H}=\left\{
\begin{array}{ll}
\frac{\rho _{0}c_{0}^{2}(1-\frac{V_{H}}{V_{0}})}{(\lambda -1)^{2}(\frac{%
\lambda }{\lambda -1}\times \frac{V_{H}}{V_{0}}-1)^{2}}, & V_{H}\leq V_{0}
\\
\rho _{0}c_{0}^{2}(\frac{V_{H}}{V_{0}}-1), & V_{H}>V_{0}%
\end{array}%
\right.
\end{equation}
In this paper, the transformation of specific internal energy $%
E-E_{H}(V_{H}) $ is taken as the plastic energy. Both the shock compression
and the plastic work cause the increasing of temperature. The increasing of
temperature from shock compression can be calculated as:
\begin{equation}
\frac{\mathrm{d}T_{H}}{\mathrm{d}V_{H}}=\frac{c_{0}^{2}\cdot \lambda
(V_{0}-V_{H})^{2}}{c_{v}\big[(\lambda -1)V_{0}-\lambda V_{H}\big]^{3}}-\frac{%
\gamma (V)}{V_{H}}T_{H}.  \label{eq-eos-temprshock}
\end{equation}%
where $c_{v}$ is the specific heat. Eq.(\ref{eq-eos-temprshock}) can be
derived from thermal equation and the Mie-Gr\"{u}neissen state of equation%
\cite{explosion}. The increasing of temperature from plastic work can be
calculated as:
\begin{equation}
\mathrm{d}T_{p}=\frac{\mathrm{d}W_{p}}{c_{v}}  \label{eq-eos-temprplastic}
\end{equation}%
Both the Eq.(\ref{eq-eos-temprshock}) and the Eq.(\ref{eq-eos-temprplastic})
can be written as the form of increment.

In this paper we choose aluminum as the sample material. The
corresponding parameters are $\rho_{0}=2700$ kg/m$^{3}$, $E=69$ Mpa,
$\nu =0.33$, $\sigma _{Y0}=120$ Mpa, $E_{\tan }=384$ MPa,
$c_{0}=5.35$ km/s, $\lambda=1.34$, $c_{v}=880$ J/(Kg$\cdot $K),
$k=237$ W/(m$\cdot $K) and $\gamma_0=1.96$ when the pressure is
below $270$ GPa. The initial temperature of the material is 300 K.

\begin{figure}[tbp]
\caption{(in JPG format) Configurations with temperature contours.
$\protect\delta=1.4$ and $v_{init}=1000$m/s. From left to right,
t=500ns,
1100ns, 1400ns, and 1700ns, respectively. The length unit here is 10 $%
\protect\mu $m. }
\end{figure}

\begin{figure}[tbp]
\centerline{\epsfig{file= 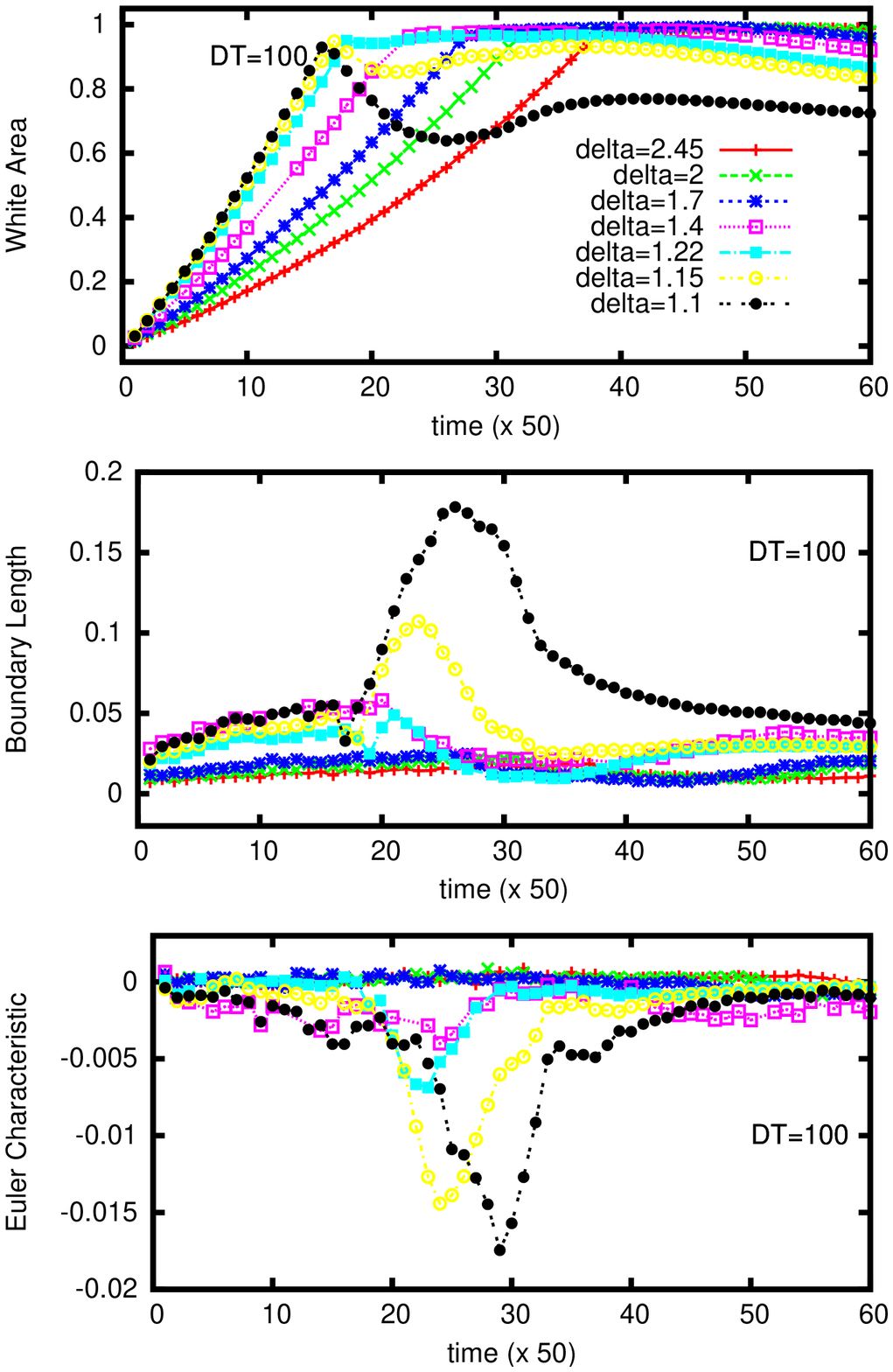, bbllx=103 pt, bblly=99
pt,bburx=509 pt,bbury=701 pt, width=0.42\textwidth,clip=}}
\caption{(Color online) Minkowski measures for  cases with various
porosities. $T_{th}=400$%
K. The values of porosity are shown in the legend.}
\end{figure}

\begin{figure}[tbp]
\centerline{\epsfig{file= 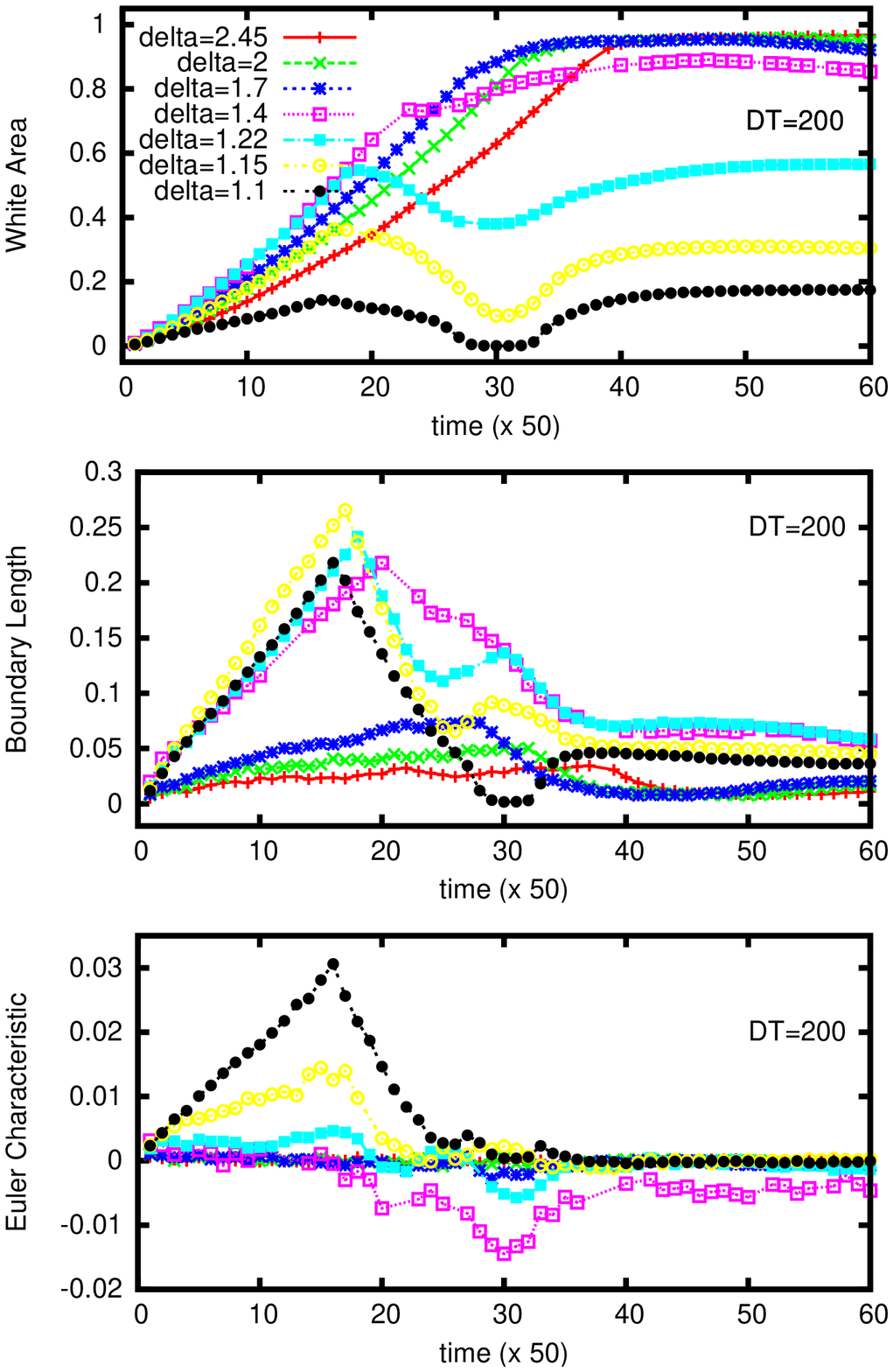, bbllx=103 pt, bblly=99
pt,bburx=509 pt,bbury=701 pt, width=0.42\textwidth,clip=}}
\caption{(Color online) Minkowski measures for  cases with various porosities.
$T_{th}$%
=500K. The values of porosity are shown in the legend.}
\end{figure}

\begin{figure}[tbp]
\centerline{\epsfig{file= 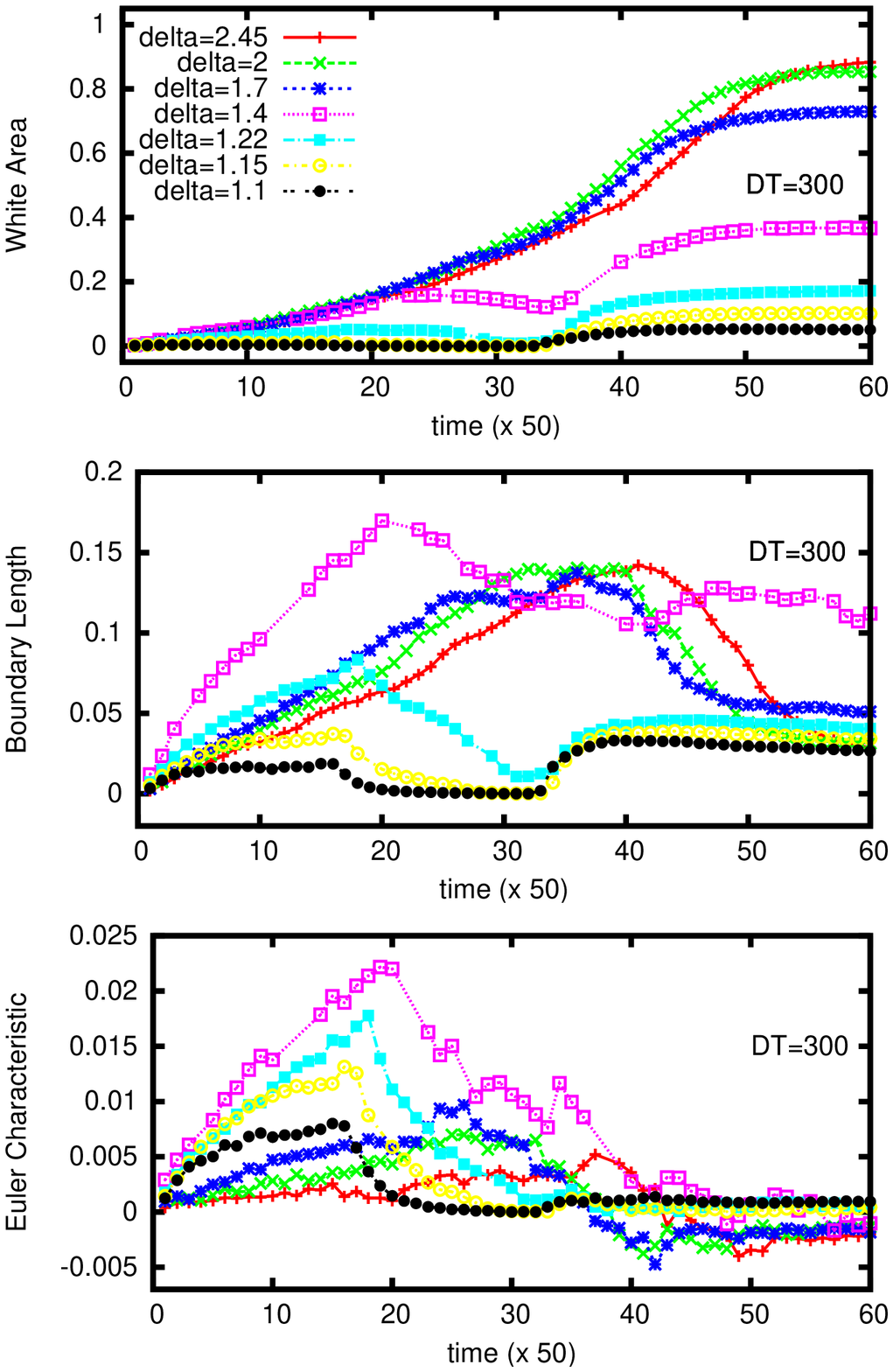, bbllx=103 pt, bblly=99
pt,bburx=509 pt,bbury=701 pt, width=0.42\textwidth,clip=}}
\caption{(Color online) Minkowski measures for cases with various porosities.
 $T_{th}$%
=600K. The values of porosity are shown in the legend.}
\end{figure}

\section{Simulation results and physical interpretation}

In our numerical experiments the porous material is fabricated by a
solid material body with an amount of voids randomly embedded. We
denote the mean density of the porous body as $\rho $ and the
density of the solid portion as $\rho _{0}$. The porosity is defined
as $\delta =\rho _{0}/\rho $. The present work concentrates on
two-dimensional case and the porosity $\delta $ is controlled by the
total number $N_{void}$ and mean size $r_{void}$ of voids embedded.
The shock wave reacting on the target porous body is loaded via a
colliding by a rigid wall with the same material. We choose the
coordinate system where the rigid wall is horizontal and keeps
static at the position $y=0$, the target porous body is on the upper
side of the rigid wall and moves towards the rigid wall at a
velocity $-v_{init}$. The porous body begins to touch the rigid wall
at the time $t=0$. The simulated porous body is initially 1 mm in
width and 5 mm in height, as shown in Fig. 1. Periodic boundary
conditions are set in the horizontal directions, which means the
real system under consideration is composed of many of the simulated
ones aligned periodically in the horizontal direction.

\begin{figure}[tbp]
\caption{(in JPG format)  Configurations with temperature contours.
$\protect\delta=1.4$ and $v_{init}=500$m/s. From left to right, t =
500 ns, 1500 ns, 2000 ns, and 2500 ns, respectively. The length unit here is 10 $%
\protect\mu $m. }
\end{figure}

\begin{figure}[tbp]
\centerline{\epsfig{file= 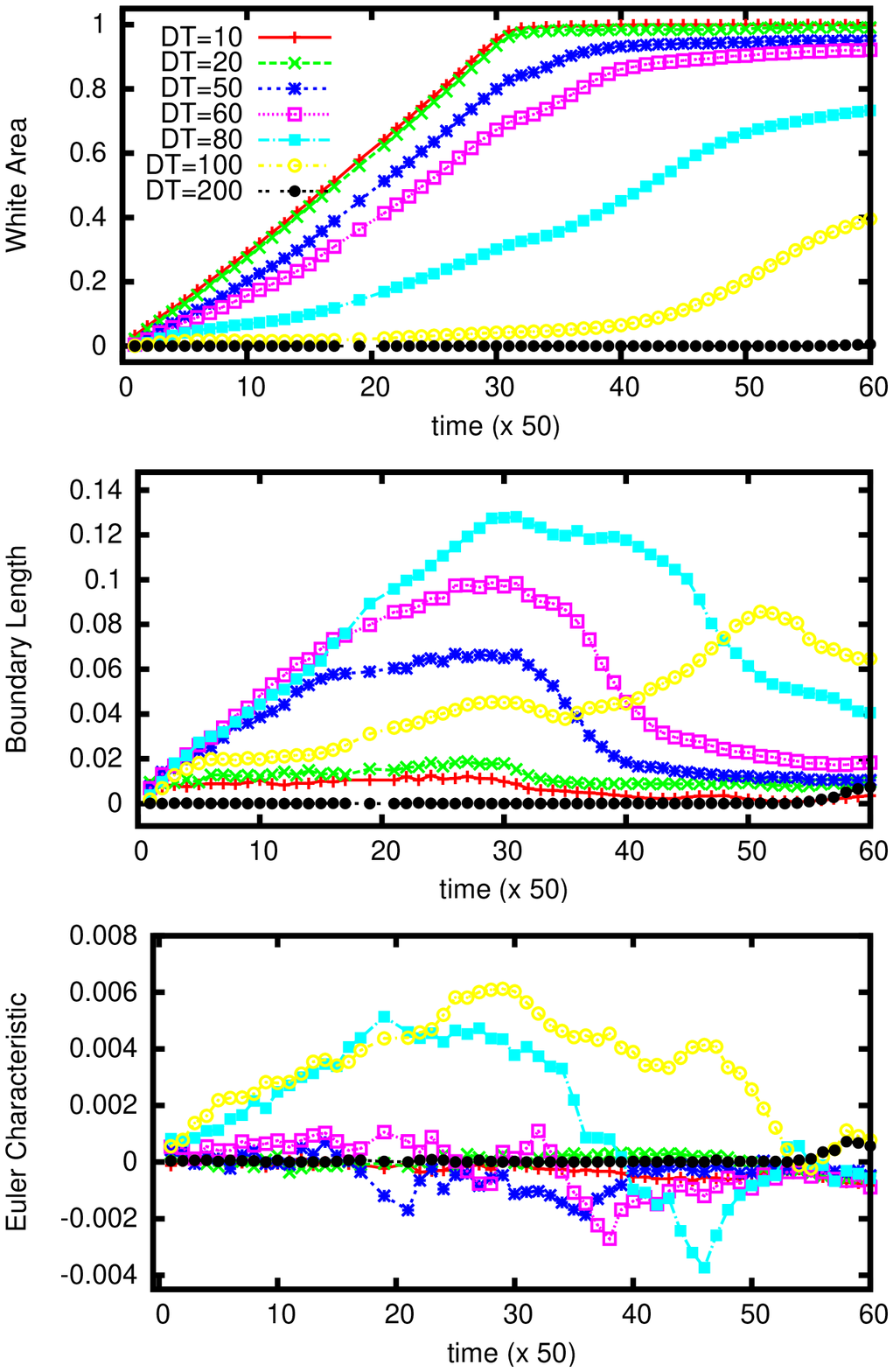, bbllx=103 pt, bblly=99
pt,bburx=509 pt,bbury=701 pt, width=0.42\textwidth,clip=}}
\caption{(Color online) Minkowski measures for the case of $\protect%
\delta=1.4$ and $v_{init}=500$m/s. The values of contour level are shown in
the legend.}
\end{figure}

\begin{figure}[tbp]
\centerline{\epsfig{file= 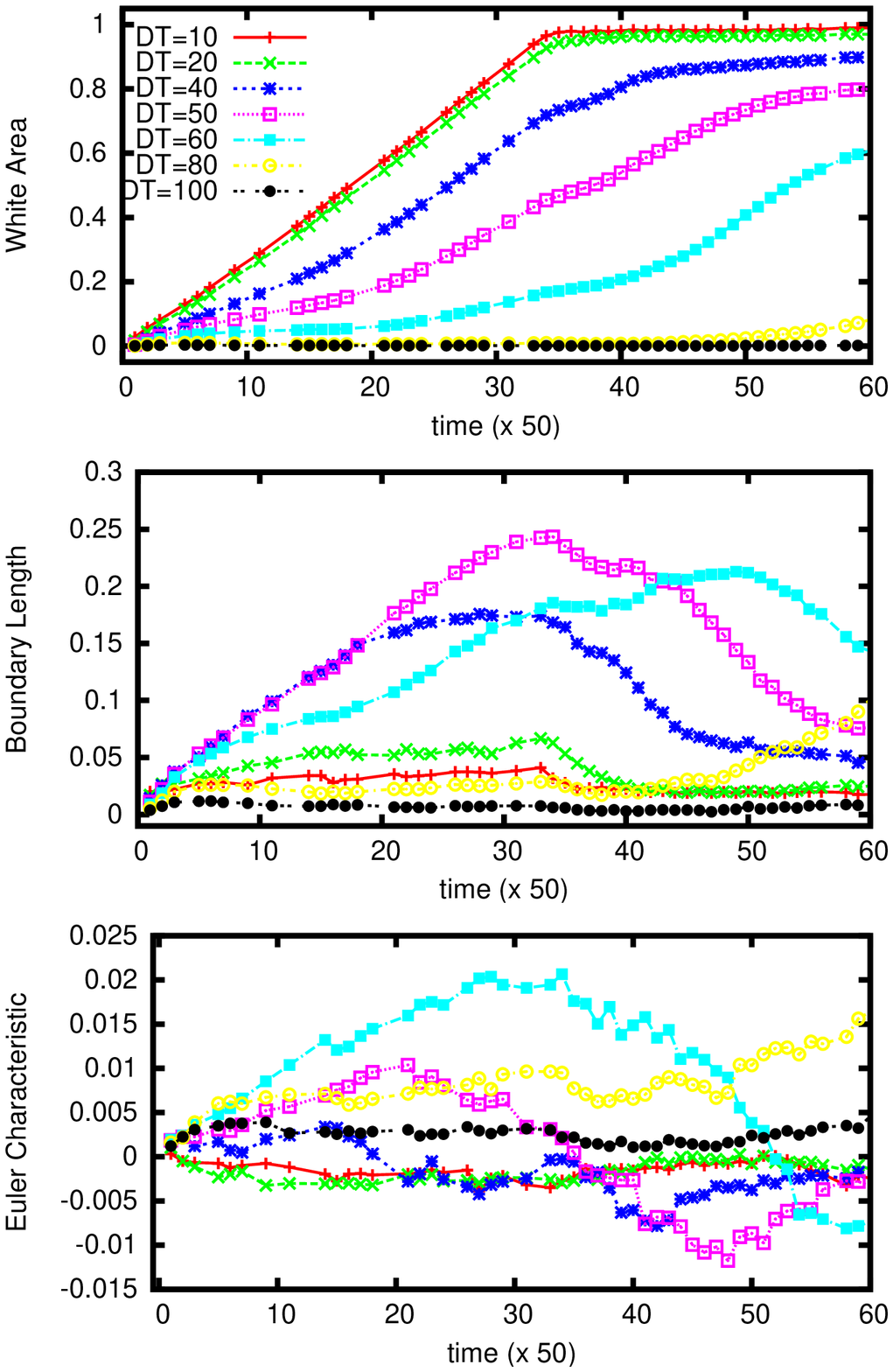, bbllx=103 pt, bblly=99
pt,bburx=509 pt,bbury=701 pt, width=0.42\textwidth,clip=}}
\caption{(Color online) Minkowski measures for the case of $\protect%
\delta=1.4$ and $v_{init}=400$m/s. The values of contour level are shown in
the legend.}
\end{figure}

\begin{figure}[tbp]
\centerline{\epsfig{file= 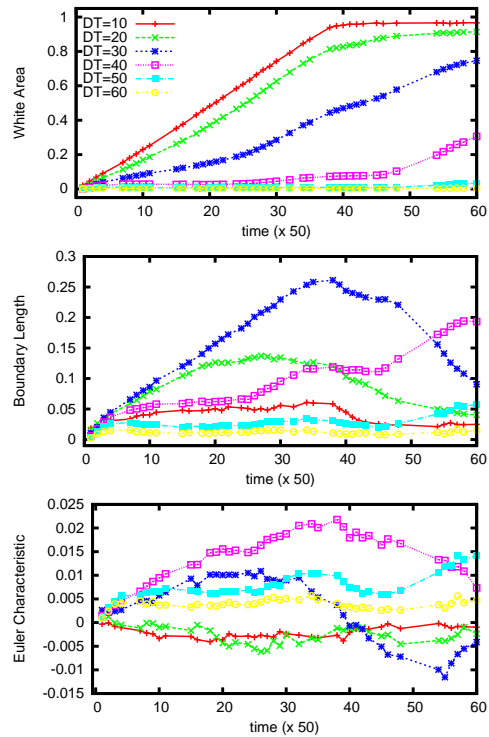, bbllx=103 pt, bblly=99
pt,bburx=509 pt,bbury=701 pt, width=0.42\textwidth,clip=}}
\caption{(Color online) Minkowski measures for the case of $\protect%
\delta=1.4$ and $v_{init}=300$m/s. The values of contour level are shown in
the legend.}
\end{figure}

\subsection{Case with $\protect\delta =2$ and $v_{init}=1000$m/s}

Figure 1 shows a set of snapshots for a procedure that a shock wave
is reacting on a porous body, where the contours denote temperature.
From blue to red, the temperature increases. The porosity $\delta
=2$, $v_{init}=1000$m/s. The time t=500ns, 1500ns, 2000ns, 2500ns
for the four snapshots from left to right. It is clear that,
different from the case with uniform material, the original shock
wave is scattered and dispersive in the porous body. The first two
snapshots show the loading procedure. When $t=500$ ns, the early
compressive
waves arrive at about $y=1$ mm; when $t=1500$ ns, they arrive at about $%
y=3.1 $ mm. The last two snapshots show the procedure of
downloading. When compressive waves arrive at the upper free
surface, rarefactive waves are reflected back into the target porous
body. Under the tension wave, the height of the porous body
increases with time. In fact, before the compressive waves arrive at
the upper free surface, a large number of local downloading
phenomena have occurred within the porous body. When the initial
shock wave or a compressive wave encounters a void, rarefactive
waves are reflected back and propagate within the compressed
portion, which destroys the original possible equilibrium state
there. Since the details of wave series are very complex, when we
mention the value of a state variable, for example the density, we
refer to its local mean value.

To perform the Minkowski functional analysis for the temperature
map, we can choose a threshold temperature $T_{th}$ and pixelize the
map into white regions (with $T\geq T_{th}$) and black regions (with
$T<T_{th}$ ). Figure 2 shows the Minkowski measures for the same
procedure as in Fig.1. \textquotedblleft $DT$ \textquotedblright\ in
the legend means $T_{th}-300$. The unit of temperature is K. The
time unit is ns. When $DT$ is very small, the wave front is nearly a
plane, which is similar to the case with shock reacting on uniform
solid material. When $DT=10$K, the total fractional white area $A$
increases up to be nearly $1$ at the time $t=1600$ ns and keeps this
value until the time $t=2600$ns, then has a slight decreasing. This
means the early compressive wave arrives at the upper free surface
at about, in fact before, the time $t=1600$ ns, nearly all material
particles in the target body have a temperature beyond $310$ K
during the following $1000$ns. In the downloading procedure the
rarefactive waves  make a very small fraction of material particles
decrease their temperature to below $310$ K.
With the increase of $DT$, the white area $A$ decreases. For the case with $%
DT=100$ K, when $t=1900$ ns, the white area arrives at a steady value $0.96$%
, which means $4\%$ of the material particles could not get a
temperature higher than $400$ K in the whole procedure shown here.
Compared with the case of $DT=10$K, we can get another piece of
information, the temperature increase in shocked portion of porous
material is much slower than in shocked uniform solid material. We
can find the physical reason for this by considering the void
effects in shocked porous body. When the compressive wave arrives at
a void, it is decomposed of many components. The components in the
solid portion move forwards more quickly, while the portion facing
the void may result in jet phenomenon. When jetted material hit the
downstream wall of the void, new compressive waves are created. At
the same time, the void reflects rarefactive wave back to the
compressed region. A large number of similar processes exist in the
shocked porous system. Thus, the shock loading procedure in the
porous body is manifested as successive reactions of many
compressive and rarefactive waves. In the shock-loading procedure,
the compressive waves dominate. Each plastic deformation makes a
temperature increment. The curve for the case of $DT=200$ K can be
interpreted in a similar way. When $DT$ increases from $200$K to
$300$K, the curve of white area has a significant variation. For the
case of $DT=400$K, the white area arrives at $0.2$ at the time
$t=3000$ns, which means $80\%$ of material particles could not get a
temperature higher than $700$ K up to this time. When $DT=500$K, the
white area keeps nearly zero during the whole procedure shown here,
which means no local temperature is higher than $800$K
in the system up to the time $t=3000$ns. For cases with $DT=300$K, $330$K, $%
360$K and $400$K, after the initial slow increasing period, the
white (hot) area has a quick increasing period. The latter indicates
that a large amount of \textquotedblleft
hot-spots\textquotedblright\ in the previously compressed region
coalesced during that period. After that the increasing of $A$ with
$t$ shows a slowing-down. The slope of the $A(t)$ curve
approximately corresponds to the mean propagation speed of some
components of the compressive waves. Therefore, the first Minkowski
measure indicates that, in porous material, when a velocity $D$ of
the compressive-wave-series mentioned, the corresponding
contour-level of a state variable like temperature should also be
stated. From this figure, it is clear that $D(T_{th})$ decreases
with the increasing of $T_{th}$; The total fractional white (hot)
area $A(t)$ shows a parabolic behavior during the initial period;
When $DT$ approaches $0$, $A(t)$ behavior goes back to be linear.

Now we go to the second Minkowski measure, the boundary length $L$.
To
understand this measure, we can consider the three-dimensional plot of $%
T(x,y)$ as a mountain. In the case where the mountain has only one
peak, when we increase the contour level $T_{th}$, the white area
$A$ decreases, and the boundary length $L$ decreases, too. But in
the case where the mountain has more than one peaks, the situation
will not be so simple: the white area $A$ may decrease while the
boundary length $L$ increases. For the case of $DT=10$K, after the
initial increase corresponding to the getting contact of the target
body with the rigid wall, the
boundary length $L$ keeps a small constant for a long time until about $t=2600$%
ns. The fact that the boundary length $L$ keeps constant while the
white area $A$ increase means also that the compressive wave is
propagating towards the upper free surface and the wave front is
nearly a plane in the pixelized temperature map. The increasing of
boundary length $L$ after the time $t=2600$ns is
companying with the decreasing of white area $A$, which means some small black (cold) spots occur.
The curves for $DT=100$ K and $%
DT=200$K show similar information. They first increase with time due
to the appearance of more \textquotedblleft
hot-spots\textquotedblright , then decreases due to the coalesce of
\textquotedblleft hot-spot\textquotedblright , finally increase,
companied by the slight decrease of
the total fractional white area $A$.
When $DT=300$K, during the period with $%
1500 $ns $<t<2500$ns, the white area $A$ increases, while the total
fractional boundary length $L$ is nearly a constant. Considering
that the wave front has not been a plane any more for this threshold
 temperature, this result indicates the following information:
 during this period, the
compressive waves propagate forwards, more scattered
\textquotedblleft hot-spots\textquotedblright\ appeare in the newly
compressed region; at the same time, some previous scattered
\textquotedblleft hot-spots\textquotedblright\ coalesce. From $2500
$ns to $3000$ns, the white area $A$ increases very slowly, but the
boundary length $L$ decreases quickly. This result show that the
increasing of white area $A$ is mainly due to coalesce of previous
scattered \textquotedblleft hot-spots\textquotedblright. The curves
for $DT=330$K and $DT=360$K can be understood in the similar way.
For the present shock strength, only very few
material particles can get a temperature beyond $700K$ before the time $%
t=2000$ns. Therefore, the boundary length $L$ for the case with
$DT=400$K has a meaningful increase only after $t=2000$ns.

When $DT$ is small, $T>T_{th}$ in (nearly) all of the compressed portion and
$T<T_{th}$ in the uncompressed part of the material body. The temperature
map shows a highly connected structure with (nearly) equal and very small
amount of black and white domains. So, the Euler characteristic $\chi $
keeps close to zero in the whole shock-loading procedure and the mean
curvature $\kappa $ is nearly zero. The value of $\chi $ decreases to be
evidently less than zero in the downloading procedure, which indicates that
the number of domains with $T<T_{th}$ increases. (See the $\chi (t)$ curves
for cases with $DT=10$, $DT=100$ and $DT=200$ in Fig.2.) With the increasing
of the contour level $T_{th}$, more regions changes their color from white ($%
T>T_{th}$) to black ($T<T_{th}$). The pattern evolution in the shock-loading
procedure can be regarded as that scattered white domains appear gradually
with time in the black background. So the Euler characteristic $\chi $ is
positive and increasing with time. (See the $\chi (t)$ curves for cases with
$DT=300$, $DT=330$ and $DT=360$ in Fig.2.) When the contour level $T_{th}$
is further increased up to $700$K, a meaningful fraction of material
particles could not get a temperature higher than the contour level $T_{th}$%
. The saturation phenomenon in the $\chi $ curve during the period, $550$ns $%
<t<2100$ ns, indicates that the numbers of connected
\textquotedblleft hot\textquotedblright\ and \textquotedblleft
cold\textquotedblright\ domains vary with time in a similar way. The
increase of $\chi $ in the period, $2100 $ns $<t<2500$ns, is due to
that the rarefactive waves make mean-temperature decrease,
correspondingly, some connected \textquotedblleft
hot-domains\textquotedblright\ are disconnected as scattered
\textquotedblleft hot-spots\textquotedblright\ again. For the case
of $DT=500$K, the pixelized temperature map is nearly in black. So,
the Euler characterization $\chi $ is nearly zero.

\subsection{Effects of porosity}

Figure 3 shows a set of snapshots for the case with a lower porosity, $%
\delta =1.4$. The other conditions are the same as in Fig.1. From
left to right, the four configurations correspond to the times,
$t=500$ns, $1100$ns, $1400$ns and $1700$ns. Compared with the
snapshots in Fig.1, it is clear that the propagation velocity of
compressive wave increases with the decreasing of porosity. At time
$t=500$ns, in the system with $\delta =1.4$, the compressive wave
arrives at about $y=1750\mu $m; while in the system with $\delta
=2$, the compressive wave only arrives at about $y=1000\mu $m. In
the case of $\delta =1.4$, the compressive wave has arrived the top
free surface and the rarefactive wave has been reflected back to the
target body before the time $t=1400$ns; while in the case of $\delta
=2$, the shock-loading procedure has not been finished up to
$t=1500$ns.

Figure 4 shows the Minkowski measures for cases with various porosities,
where $T_{th}=400$K and the values of porosity, $\delta =2.45$, $2$, $1.7$, $%
1.4$, $1.22$, $1.15$, $1.1$ are shown in the legend. In the
subfigure for white area $A$, the initial shock-loading part
presents meaningful information: the velocity $D$ of
the compressive-wave-series is smaller for a higher porosity $\delta $%
. The most significant property in the subfigure for boundary length
$L$ is
that the largest boundary length $L_{max}$ increases as $\delta $ decreases. When $%
\delta =1.1$, the total boundary length $L$ gets the maximum value at about $%
t=1250$ns. This result indicates that the highest temperature in
shocked porous material decreases when the porosity approaches $1$.
The Euler characteristic $\chi $ becomes more negative when the
porosity $\delta $ decreases from $2.45$ to $1.1$, which means the
disconnected \textquotedblleft cold\textquotedblright\ domains with
$T<400$K dominate more the image.

Figures 5 and 6 show the Minkowski measures for the same porosities but
higher temperature thresholds, $T_{th}=500$K and $T_{th}=600$K. They present
supplementary information to Fig. 4. For cases with $\delta =1.4$, $1.22$, $%
1.15$ and $1.1$, only $88\%$, $55\%$, $36\%$ and $15\%$ of the material
particles get the temperature higher than $500$K. For cases with $\delta =1.4
$ and $1.22$, and only $16\%$ and $6\%$ get the temperature higher than $600$%
K in the shock-loading procedure. When $T_{th}=500$K, the case with
$\delta =1.15$ has the maximum boundary length and the case with
$\delta =1.1$ has the maximum Euler characteristic. When
$T_{th}=600$K, the case with $\delta =1.4$ has the maximum boundary
length and maximum Euler characteristic, which means the
\textquotedblleft hot-spots\textquotedblright\ with $T>600$K are
scatteredly distributed in the \textquotedblleft
cold\textquotedblright\ background with $T<600$K.

\subsection{Effects of initial shock-wave-strength}

\begin{figure}[tbp]
\centerline{\epsfig{file= 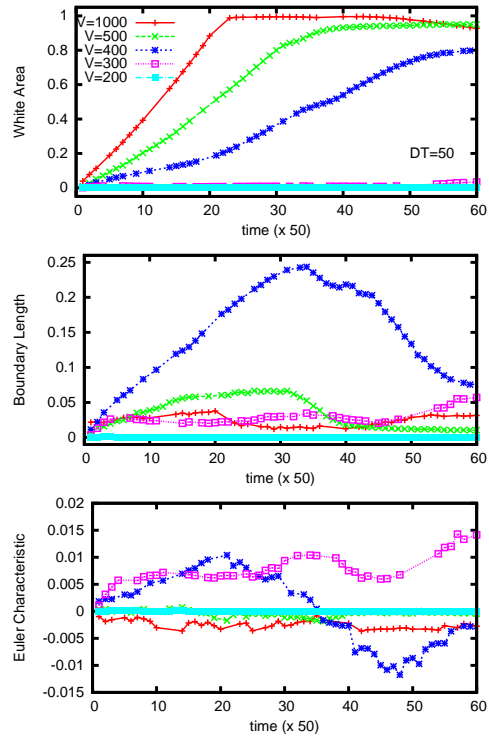, bbllx=103 pt, bblly=99
pt,bburx=509 pt,bbury=701 pt, width=0.42\textwidth,clip=}}
\caption{(Color online) Minkowski measures for cases with various shock strengths. $\protect%
\delta=1.4$. The values of initial impacting speed $v_{init}$ are shown in
the legend.}
\end{figure}

We now study the effects of different initial impacting speeds. Figure 7
shows a set of snapshots for the case with $\delta =1.4$ and $v_{init}=500$%
m/s. From left to right, the four configurations are for the times $t=500$%
ns, $1500$ns, $2000$ns and $2500$ns. From the first two, we observe
the upward propagation of compressive wave in the target body. From
the last two, we observe the downward rarefactive effects. Compared
with Fig.3, it is
clear that the velocity $D$ of compressive-wave-series and the highest temperature $%
T_{\max }$ decreased. The Minkowski meansures for this procedure is
shown in Fig. 8. Such a shocking procedure could not produce
\textquotedblleft hot-spot\textquotedblright\ with $T=500$K.
High-temperature \textquotedblleft Hot-area\textquotedblright\
continue to increase even after some precursory compressive waves
have scanned all the target body and some rarefactive waves have
come into the target body from the upper free surface. Up to the
time $t=3000$ns, the
fractional area of \textquotedblleft Hot-spots\textquotedblright\ with $%
T>400 $K reaches $40\%$, the fractional area for $T>380$K reaches
$74\%$, that for $T>360$K reaches $91\%$. The contour-level with
$T=380K$ has the largest boundary length at about $t=1500$ns when
 the \textquotedblleft hot-spots\textquotedblright\ mainly distribute
scatteredly in the \textquotedblleft cold\textquotedblright\
background. Figures 9 and 10 show the Minkowski measures for cases
with the same porosity but lower initial impacting speeds.
$v_{init}=400$m/s in Fig.9 and $v_{init}=300$m/s in
Fig.10. With the decrease of initial impact speed, the highest temperature $%
T_{\max }$ in the system further decreases; the total fractional white area $%
A$ for low contour-level, for example $DT=10$K, increases with time in a
more linear way.

We compare Minkowski measures for different initial impacting speeds in Fig.
11, where $\delta =1.4$, $DT=50$K, $v_{init}=1000$ms, $500$m/s, $400$m/s, $%
300$m/s, and $200$m/s. It is clear that the higher the initial impacting
speed, the closer to be linear the $A(t)$ curve. The case of $%
v_{init}=400$m/s has the longest total boundary separating the
\textquotedblleft hot\textquotedblright\ and \textquotedblleft
cold\textquotedblright\ domains. For this case, disconnected
\textquotedblleft hot\textquotedblright\ regions dominate the image
from the topology side in the shock-loading procedure; disconnected
\textquotedblleft cold\textquotedblright\ regions dominate in the
downloading procedure.

\section{Conclusions}

Under shock wave reaction, the porous material is globally in a
nonequilibrium state and shows complex dissipative structures. We
pixelize the map of temperature into Turing patterns and introduce
morphological measures for it.
 Relevance of the total fractional white area $A$,
boundary length $L$ and the Euler characteristic $\chi$ to the
thermodynamical properties of material is revealed.  Various
experimental conditions are simulated via the material-point method.
Numerical results indicate that, the shock wave reaction results in
a complicated sequence of compressions and rarefactions in porous
material. The increasing rate of $A$ roughly gives the velocity $D$
of a compressive-wave-series. When a velocity $D$ is mentioned, the
corresponding threshold contour-level of the temperature should also
be stated. When the threshold contour-level increases, $D$ becomes
smaller. The area $A$ increases parabolically with time $t$ during
the initial period. The $A(t)$ curve goes back to be linear in the
following three cases: (i) when the porosity $\delta$ approaches 1,
(ii) when the initial shock becomes stronger, (iii) when the
contour-level approaches the minimum value of the temperature. The
area with high-temperature may continue to increase even after the
early compressive-waves have arrived at the downstream free surface
and some rarefactive-waves have come back into the target body. In
the case of energetic material needing a higher temperature for
initiation, a higher porosity is preferred and the material may be
initiated after the precursory compressive-waves have scanned all
the target body.  One may desire the fabrication of a porous body
and choose the appropriate shock strength according to what needed
is scattered or connected hot-spots. The same measures can also be
used to analyze the maps of other physical variables, like the
density, velocity, or various stresses. With the Minkowski measures,
the dependence on experimental conditions is reflected simply by a
few coefficients. They may be used as order parameters to classify
the maps of state variable in a similar way like thermodynamic phase
transitions.

\ack{ We warmly thank Jianguo Wang, Hua Li, Yangjun Ying for helpful
discussions on shock waves and porous material. A.Xu is grateful to
Drs. G. Gonnella and A. Lamura for constructive discussions on
Minkowski functionals. This work is supported by Science Foundations
of LCP and CAEP, national Science Foundation of China (under Grant
Nos. 10702010,10775018 and 10604010).


\end{document}